\documentclass[letterpaper, 10 pt, conference]{ieeeconf}  % Comment this line out if you need a4paper

\IEEEoverridecommandlockouts                              % This command is only needed if 
\IEEEoverridecommandlockouts
\usepackage{cite,booktabs,makecell,multirow,pifont}
\usepackage{amsmath,amssymb,amsfonts}
\usepackage{algorithmic}
\usepackage{graphicx}
\usepackage{textcomp}
\usepackage[hidelinks]{hyperref} 
\usepackage[caption=false, labelformat=parens, subrefformat=parens]{subfig}
\usepackage{xcolor}
\newcolumntype{?}{!{\vrule width 1pt}}
\title{\LARGE \bf 
Recover from Horcrux: A Spectrogram Augmentation Method for Cardiac Feature Monitoring from Radar Signal Components}

\author{Yuanyuan~Zhang$^{1}$, Sijie~Xiong$^{2}$, Rui~Yang$^{1,*}$, Eng~Gee~Lim$^{1}$, Yutao~Yue$^{3,*}$% <-this % stops a space
\thanks{This research has been approved by the University Ethics Committee of Xi'an Jiaotong-Liverpool University with proposal number ER-SAT-0010000090020220906151929, and is partially supported by Suzhou Science and Technology Programme (SYG202106), Jiangsu Industrial Technology Research Institute (JITRI) and Wuxi National Hi-Tech District (WND).}
\thanks{$^{1}$Yuanyuan Zhang, Rui Yang and Eng Gee Lim are with the School of Advanced Technology, Xi'an Jiaotong-Liverpool University, Suzhou, 215123, China {\tt\small Yuanyuan.Zhang16@student.xjtlu.edu.cn; R.Yang@xjtlu.edu.cn; Enggee.Lim@xjtlu.edu.cn}}%
\thanks{$^{2}$Sijie Xiong is with the Faculty of Information Science and Electrical Engineering, Kyushu University, 819-0395, Fukuoka, Japan {\tt\small xiong.sijie.630@s.kyushu-u.ac.jp}}%
\thanks{$^{3}$Yutao Yue is with the Thrust of Artificial Intelligence and Thrust of Intelligent Transportation, The Hong Kong University of Science and Technology (Guangzhou), Guangzhou 511400, China, and also with the Institute of Deep Perception Technology, JITRI, Wuxi 214000, China {\tt\small yutaoyue@hkust-gz.edu.cn}}%
\thanks{\textsuperscript{*}\textit{Corresponding authors: Rui Yang and Yutao Yue.}}
\thanks{Submitted to IEEE EMBC 2025}
}

\begin{document}
\maketitle
\thispagestyle{empty}
\pagestyle{empty}

\begin{abstract}
Radar-based wellness monitoring is becoming an effective measurement to provide accurate vital signs in a contactless manner, but data scarcity retards the related research on deep-learning-based methods. Data augmentation is commonly used to enrich the dataset by modifying the existing data, but most augmentation techniques can only couple with classification tasks. To enable the augmentation for regression tasks, this research proposes a spectrogram augmentation method, Horcrux, for radar-based cardiac feature monitoring (e.g., heartbeat detection, electrocardiogram reconstruction) with both classification and regression tasks involved. The proposed method is designed to increase the diversity of input samples while the augmented spectrogram is still faithful to the original ground truth vital sign. In addition, Horcrux proposes to inject zero values in specific areas to enhance the awareness of the deep learning model on subtle cardiac features, improving the performance for the limited dataset. Experimental result shows that Horcrux achieves an overall improvement of $16.20\%$ in cardiac monitoring and has the potential to be extended to other spectrogram-based tasks. The code will be released upon publication.

\end{abstract}

\section{Introduction}
Contactless vital sign monitoring is important for various applications in future scenarios such as smart home and in-cabin monitoring, because it is impossible or inconvenient to access the human body for certain daily monitoring cases~\cite{zhang2024radarODE,chen2022contactless}. Among all kinds of contactless-based sensors (e.g., camera, acoustic sensor, Wi-Fi), radar can perform unobtrusive monitoring with high accuracy in the presence of light or sound noises and has been widely investigated for monitoring vital signs such as heart rate (HR), seismocardiography (SCG), electrocardiogram (ECG)~\cite{zhang2024radarODE,chen2022contactless,ha2020contactless}. 

Two paradigms of methods are commonly used in recovering vital signs from radar signals, i.e., traditional signal-processing method and deep learning-based method~\cite{zhang2023overview}. The methods in the first paradigm are normally based on certain intrinsic features of cardiac activities (e.g., periodicity) and contain many well-known algorithms such as independent component analysis (ICA) and empirical mode decomposition (EMD)~\cite{zhang2023overview}, while the recent research also turns to investigate the deep learning-based method for yielding fine-grained cardiac features (e.g, ECG), because the cardiac mechanical activities sensed by radar cannot be transformed in an explicit manner to the ECG signal as cardiac electrical activities~\cite{zhang2024radarODE}. Therefore, most studies leverage the great non-linear mapping ability of the deep learning model to learn the transformation from the numerous radar/ECG pairs.  

\begin{figure}[tb]
        \centering
        \subfloat[]{\label{fig:spec_trad}\includegraphics[width=0.5\columnwidth]{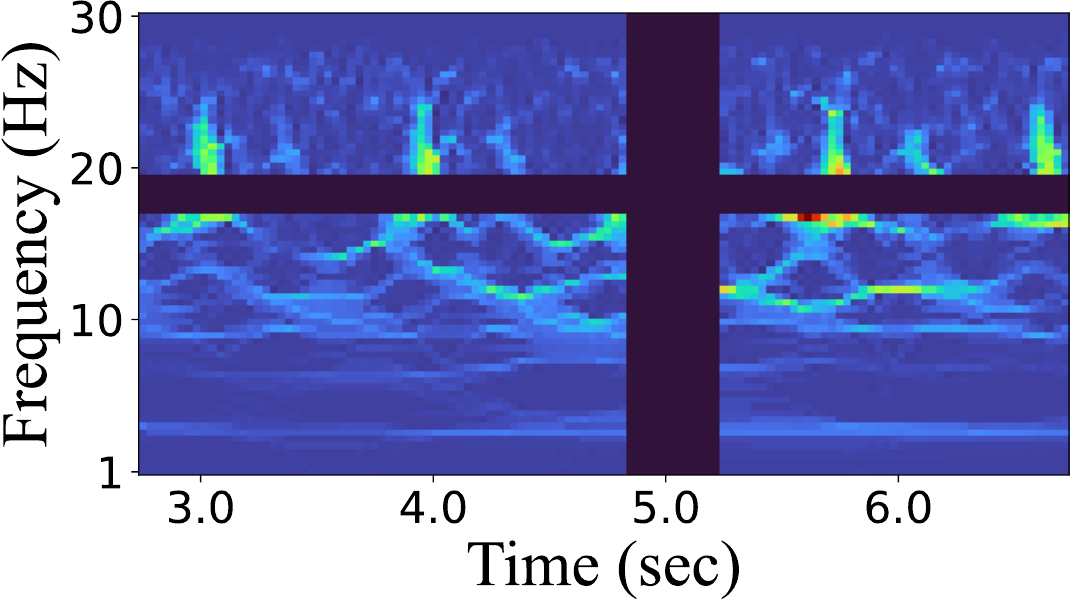}}
        \subfloat[]{\label{fig:bad}\includegraphics[width=0.5\columnwidth]{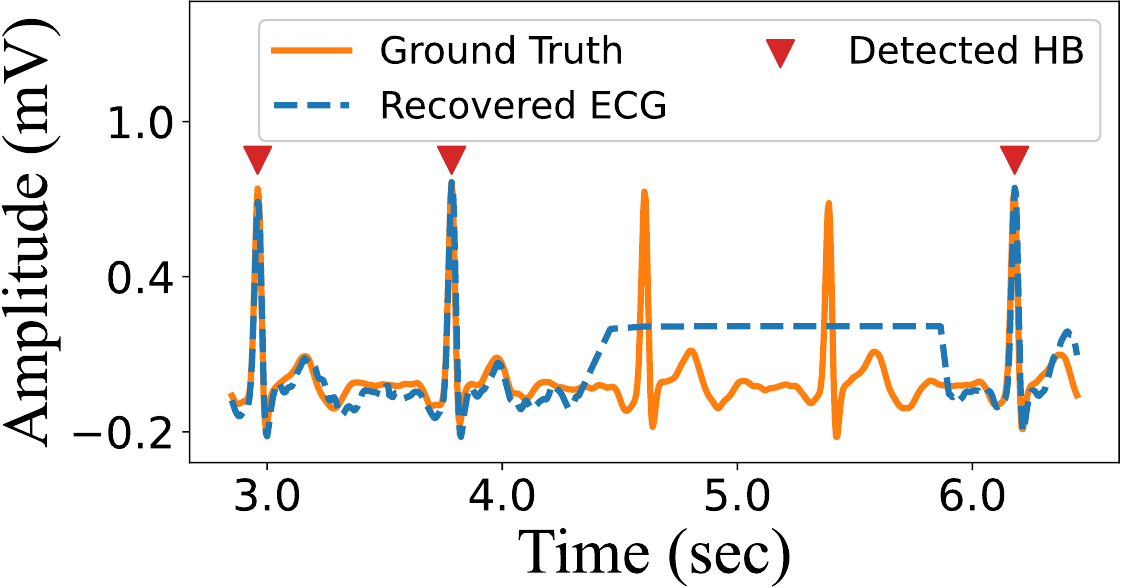}} \\
        \subfloat[]{\label{fig:spec_hp}\includegraphics[width=0.5\columnwidth]{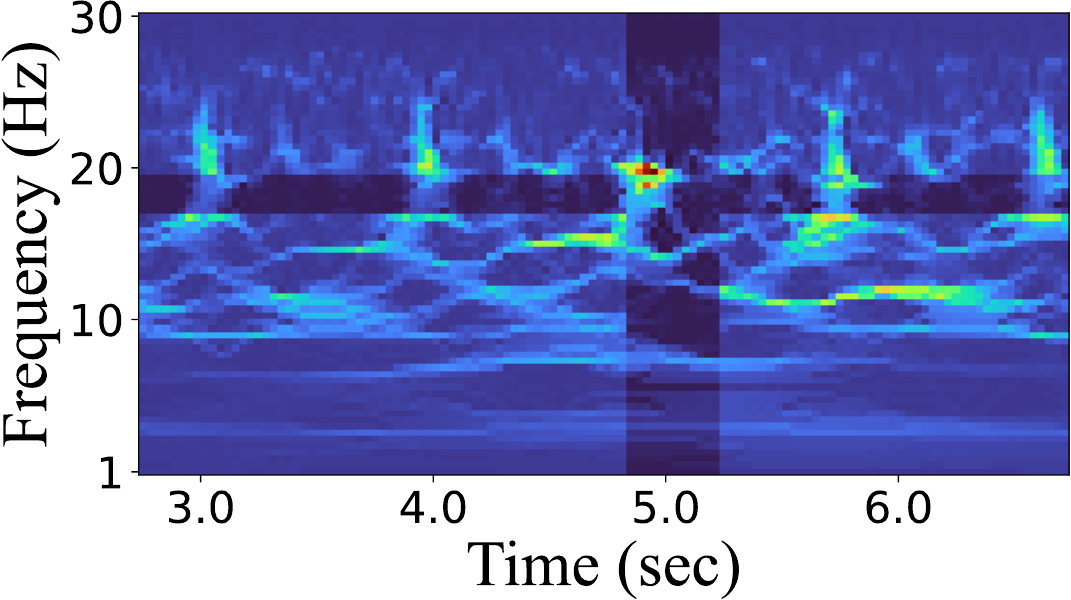}}
        \subfloat[]{\label{fig:good}\includegraphics[width=0.5\columnwidth]{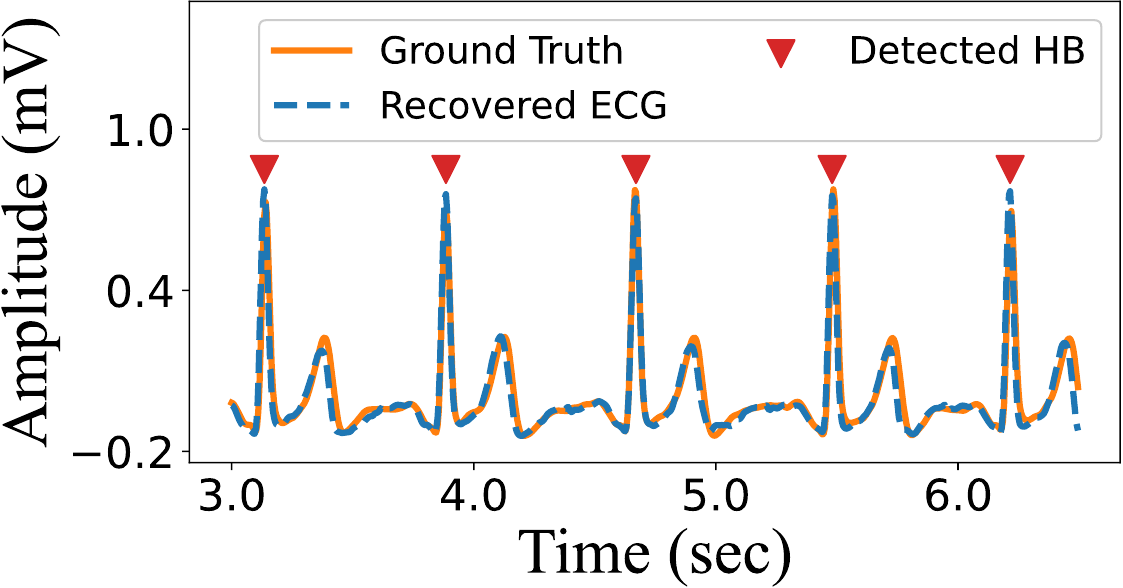}}
        \caption{Illustration of Horcrux: (a) Traditional augmentation with zero mask; (b) Missed detected heartbeat and distorted ECG recovery; (c) Horcrux with time consistency preserved; (d) Ideal cardiac monitoring with high fidelity.}
        \label{fig:hp_compare}
\end{figure}

Similar to all the deep-learning-related areas, data scarcity is an inevitable issue, especially for the radar system that requires expertise in configuration setting and signal pre-processing, and data augmentation is a popular solution by modifying the original data to enhance the diversity of dataset, improving the performance with limited datasets and the generalization for out-of-distribution (OOD) data~\cite{qin2023spatial,daniel2023between}. Most existing augmentation techniques are designed for classification problems, and the inputs (signal or image) can be safely masked/recombined/flipped/stretched without changing the ground truth (labeled class). However, the aforementioned techniques are not applicable to the regression-based vital sign reconstruction (e.g. radar-based ECG recovery), because the changes on the input cannot be directly applied on the ground truth ECG, e.g., the real ECG peaks have a similar width, therefore randomly stretching the ECG signals is unfaithful to the biological facts~\cite{swift2021stop,zhang2024radarODE}.

Different from the augmentations for classification, the regression tasks ask for a proper reorganization of the ground truth, and the selected materials (original data) should be homogeneous to restrict the potential distribution shift introduced into the augmented dataset~\cite{schneider2024anchor,li2024bifrost,yan2024predicting,zhang2024stability}. One category of the augmentation methods for regression tasks is based on mix-up techniques evolved from time-series forecasting tasks by recombining the original data within the homogeneous group and hence could generate faithful ground truth series. However, in the scope of radar-based ECG recovery, the homogeneous pairs tend to be consecutive cardiac cycles with identical features~\cite{zhang2024radarODE}, and the diversity of the dataset is hardly enhanced after the mix-up process. 

Zero mask is another augmentation technique applicable to regression tasks~\cite{he2022masked,chen2024tfpred,qin2023spatial}, because it encourages the model to focus on the most informative features in the data~\cite{he2022masked} and also leverage the unmasked contextual information in the inference stage. However, some regression tasks are highly reliant on the time-domain consistency, and directly erasing all the information is equivalent to suffering strong noises (e.g., random body movement) and can destroy the inferred output as reported by many cardiac monitoring studies~\cite{chen2022contactless,zhang2024radarODE}. 

In the literature, no augmentation method is designed for radar-based cardiac monitoring, especially considering the spectrogram as the input. Therefore, this study proposes Horcrux\footnote{In Harry Potter book series, Horcrux is an object in which wizards conceal fragments of their soul in order to become immortal.} as an augmentation method with the main contributions summarized as follows:
\begin{itemize}
\item To the best of our knowledge, Horcrux is the first augmentation method designed for extracting cardiac features from radar by preserving the intrinsic time consistency hidden in the input radar spectrogram to restrict the potential distribution shift as illustrated in Figure~\ref{fig:hp_compare}, expanding the diversity of the limited training dataset without distorting the key features.
\item To further enhance the features for vital signs, Horcrux proposes a dynamic template matching (DTM) module to adaptively identify the crucial regions and enhance the awareness of the deep learning model with respect to specific areas, encouraging the model to only focus on the domain-preserved features.
\item The experimental results illustrate that the proposed Horcrux outperforms the existing augmentation methods in both classification and regression tasks (i.e., heartbeat detection and ECG recovery), and the related cardiac features are further enhanced by DTW revealed by the ablation studies.
\end{itemize}

\section{Theoretical Background}
\subsection{Signal Model for Radar Signal}
In radar-based vital sign monitoring, the ideal sensed chest region displacement $d(t)$ normally involved with multiple vital signs as:
\begin{equation}
d(t) = d_c(t) + d_r(t)
\end{equation}
with $d_c(t)$ and $d_r(t)$ indicating cardiac and respiratory displacement, respectively~\cite{chen2021movi}.
% the phase term of the transmitted radar signal will be  modulated in terms of the distance between human body and radar and the chest region displacement $d(t)$~\cite{zhang2023overview}, and numerous techniques have been invented to unwrap the displacement $d(t)$ from the phase variation of the signal with wavelength $\lambda$ as: 
% \begin{equation}\label{equ:phase}
%  d(t) = \frac{\Delta\phi(t)\lambda }{4\pi}
% \end{equation}

Considering the recent advances in millimeter-wave radar platforms with multi-input multi-output (MIMO) technique, the respiration term can be easily eliminated~\cite{giordano2022survey,domnik2022radar} and the fine-grained cardiac activities can be further modeled as two prominent vibrations induced by aortic valve opening and closure (AO and AC)~\cite{swift2021stop,zhang2024radarODE}. The resultant $\tilde{d}(t)$ after respiration removal for $K$ cardiac cycles can be expressed as:
\begin{equation}\label{equ:vib_long}
\tilde{d}(t) = \sum_{k=1}^{K} v^k_1(t) + \sum_{k=1}^{K} v^k_2(t)
\end{equation}
with
\begin{equation}\label{equ:t1}
\begin{aligned}
 v^k_1 (t) &= \mathrm{a}^k_1 \mathrm{cos}(2\pi f^k_1 t)\exp \left(-\frac{(t-T^k_1)^2}{{b^k_1}^2}\right)\\
v^k_2 (t) &= \mathrm{a}^k_2 \mathrm{cos}(2\pi f^k_2 t) \exp \left(-\frac{(t-T^k_2)^2}{{b^k_1}^2}\right)
 \end{aligned}
\end{equation}
where $T^k_1, T^k_2$ represent the time indices of the vibrations with the central frequencies $f^k_1$, $f^k_2$ for $k^{th}$ cardiac cycle, and the rest of the parameters jointly determine the amplitude and width of two vibrations.
\begin{figure*}[tb] 
    \centering 
    \includegraphics[width=1.9\columnwidth]{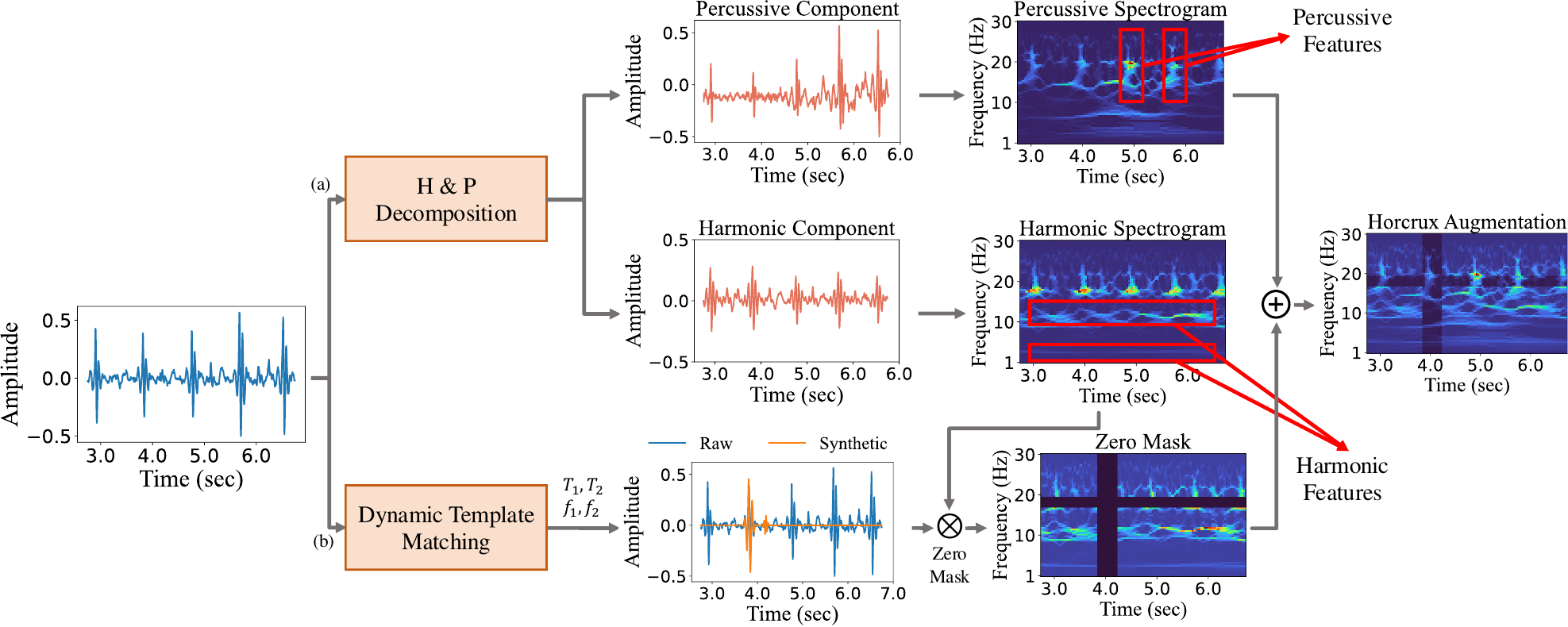}
    \caption{Pipeline of Horcrux with two branches: (a) Harmonic and percussive (H$\&$P) components decomposition; (b) Dynamic template matching (DTM).} 
    \label{fig:HP_pipline} 
\end{figure*}
\subsection{Task Definition for Cardiac Feature Extraction}
Cardiac monitoring can be generally categorized into extracting coarse and fine-grained features, with the coarse features normally based on heartbeat detection as a classification problem and fine-grained features formed as a regression task to recover certain signals (e.g., ECG, SCG). In this paper, heartbeat detection and ECG recovery are selected as the monitored cardiac features with the following definitions:
\begin{itemize}
\item \textbf{Heartbeat Detection}: The detection of heartbeat is equivalent to finding a set of time indices $\mathbf{B}$ for the most prominent vibration within each cardiac cycle as $\mathbf{B} = \{T^1_1,T^2_1,\cdots,T^K_1\}$.
\item \textbf{ECG Reconstruction}: The reconstruction of the ECG signal can be formulated as learning a mapping function $\mathcal{T}(\cdot)$
to transform the cardiac mechanical activities sensed by radar to the cardiac electrical activities described by ECG as $d_{ecg}(t) = \mathcal{T}(\tilde{d}(t))$.
\end{itemize}

\section{Methodology}
\subsection{Overview}
The overview of the proposed Horcrux is depicted in Figure~\ref{fig:HP_pipline} with two parallel modules:
\begin{itemize}
\item The H\&P decomposition module decomposes the original radar signal into the harmonic and percussive components, with the former referring to the stable component along certain frequency bands on a spectrogram and the latter representing the transient component with a pulsatile nature occurring at certain time indices, as shown in the red boxes in Figure~\ref{fig:HP_pipline}.
\item The DTM module aims to figure out the most characteristic vibration based on the signal template in (\ref{equ:vib_long}) and output the corresponding time indices $T_1, T_2$ and central frequencies $f_1,f_2$ (omit $k$ for simplicity), with the comparison between the raw and synthetic radar signal shown in Figure~\ref{fig:HP_pipline}.
\end{itemize}
The traditional augmentation methods either cannot be directly applied to the ground truth data or could potentially erase the crucial part for prediction (especially for heartbeat detection), as shown in Figure~\subref*{fig:spec_trad}. Therefore, Horcrux only masks the spectrogram of harmonic component and preserves the percussive feature, as shown in Figure~\subref*{fig:spec_hp}. In addition, by deliberately masking the related part based on $T_1, T_2, f_1, f_2$, it is expected that the well-trained deep learning model will concentrate more on the regions that help cardiac feature extraction instead of being distracted by the noises spread on the spectrogram.

\subsection{Harmonic and Percussive Components Decomposition} 
Signal decomposition is widely used in signal processing to identify components with different features and help the downstream algorithm extract target information e.g., ICA decomposes the signal from different sources based on statistical independence, and EMD decomposes the signal based on the oscillatory modes with different time scales~\cite{zhang2023overview}. Considering the nature of the cardiac activities, the radar signal reflected from the chest region reveals two characteristics on the spectrogram: (a) the stable components along certain frequency bands that represent either the heart rate frequency (near $1$Hz) or the center frequency of the prominent vibrations ($10-25$ Hz)~\cite{zhang2024radarODE}; (b) the transient components simply indicates the pulsatile heartbeats with large energy. In this research, these two components are named harmonic and percussive components (features) as highlighted in the red boxes in Figure~\ref{fig:HP_pipline}.

The decomposition process is based on the spectrogram $\mathcal{Y}(m, n)$ obtained using any time-frequency representation tools such as synchrosqueezed wavelet transform~\cite{zhang2024radarODE}, with $m$ and $n$ indicating the time frame index and frequency bin index, respectively. Then, the median filter $\mathcal{F}(\cdot)$ is applied along the time axis of $\mathcal{Y}(m,n)$ to get the harmonic-enhanced spectrogram as:
\begin{equation}
Y_h(m, n)=\mathcal{F}(\mathcal{Y}(m, n), k_h)
\end{equation}
where the median filter replaces the middle value within a window length $k_h$ by the median value in this window and is widely used in image and signal processing~\cite{saad2022addressing,mishiba2023fast}. 

Similarly, the percussion-enhanced spectrogram can be obtained using a median filter with window length $k_p$ along the frequency axis as:
\begin{equation}
Y_p(m, n)=\mathcal{F}(\mathcal{Y}(m, n), k_p)
\end{equation}

Then, a soft mask based on the Wiener filter is suggested in~\cite{fitzgerald2010harmonic} to adaptively suppress the unwanted components, and the Wiener masks $M_h$ and $M_p$ for isolating harmonic and percussive component from the original spectrogram are calculated as:
\begin{equation}
\begin{aligned}
M_h(m,n)&=\frac{Y_h(m,n)}{Y_h(m,n)+Y_p(m,n)}\\
M_p(m,n)&=\frac{Y_p(m,n)}{Y_h(m,n)+Y_p(m,n)}
\end{aligned}
\end{equation}

Lastly, the harmonic spectrogram $\mathcal{Y}_h(m, n)$ and percussive spectrogram $\mathcal{Y}_p(m, n)$ can be obtained as:
\begin{equation}
\begin{aligned}
\mathcal{Y}_h(m, n)=M_h(m, n) \otimes \mathcal{Y}(m, n)\\
\mathcal{Y}_p(m, n)=M_p(m, n) \otimes \mathcal{Y}(m, n)
\end{aligned}
\end{equation}
with $\otimes$ representing element-wise multiplication, and the resultant spectrograms are shown in Figure~\ref{fig:HP_pipline} with the corresponding features enhanced. In addition, it can be observed that certain percussive features are still revealed on the harmonic component and vice versa, because the soft Wiener mask can not totally eliminate the other component with much larger energy on the spectrogram. In addition, the Horcrux augmentation does not require a complete separation of H\&P component, H\&P decomposition module is to preserve the percussive features, therefore it is acceptable

\subsection{Dynamic Template Matching (DTM)}
The DTM module is designed to figure out the regions in the time- and frequency-domain that are closely related to the prominent vibrations, encouraging the deep learning model to focus on these regions for capturing crucial information. According to the literature~\cite{zhang2024radarODE}, the time indices $T_1,T_2$ of the prominent vibrations can be identified by the deep learning model based on their central frequencies $f_1,f_2$, and Horcrux proposes to dynamically obtain these four parameters for each input segment as a template matching problem to find the optimal parameter set:
\begin{equation}\label{equ:opt_equ}
\begin{aligned}
{\theta}^*=\underset{{\theta=\{T_1,T_2,f_1,f_2\}}}{\arg \min } &\left\|y(t)-\tilde{d}(t, \theta)\right\|_2  \\
\text{Subject to: } &T_2-T_1<\tau
\end{aligned} 
\end{equation}
where $y(t)$ is the segment from dataset and $\tilde{d}(t,\theta)$ represents the signal synthesized from (\ref{equ:vib_long}) with the parameter set ${\theta}$. The constraint represents that $T_2$ always follows $T_1$ within a distance $\tau$, because the identified vibrations should belong to the same cardiac cycle and the time interval between aortic valve opening ($T_1$) and closure ($T_2$) should be less than $\tau$ sec~\cite{swift2021stop}. 

To simplify the optimization, only one cardiac cycle is considered for one segment, and only the key parameters (i.e., $\theta=\{T_1,T_2,f_1, f_2\}$) are left to be determined, while the values of other parameters are empirically assigned as shown in Table~\ref{tab:sm_param}. In addition, the optimization problem in (\ref{equ:opt_equ}) is solved by sequential least squares programming (SLSQP) to yield optimal parameter set $\theta^*$ that indicates the time indices and central frequencies of the prominent vibrations as shown as the synthetic signal in Figure~\ref{fig:HP_pipline}.

\begin{table}[tb]
\centering
\caption{Parameters for Signal Model}
    \begin{tabular}{cc?cc?cc?cc}
    \toprule
    \textbf{Par.} & \textbf{Value} & \textbf{Par.} & \textbf{Value} & \textbf{Par.} & \textbf{Value} & \textbf{Par.} & \textbf{Value} \\
    \toprule
    $T_1$ & $0.4$  & $T_2$  & $0.85$ &$f_1$ & $10$  & $f_2$  & $23$  \\
    \midrule
    $a_1$ & $0.5$  & $a_2$  & $0.1$ & $b_1$ & $0.05$  & $b_2$  & $0.03$ \\
    \bottomrule
    \end{tabular}%
\label{tab:sm_param}
\end{table}%
\subsection{Augmented Spectrogram Generation}
The last step of Horcrux is to randomly select one vibration from $v_1,v_2$ and mask the corresponding regions on the harmonic spectrogram based on $\theta^*$ with the width of $w_t$ or $w_f$ for time or frequency domain. The applied zero mask alters the data distribution of the original spectrogram to enhance the diversity of the limited dataset, and the deep learning model will concentrate more on the masked regions selected by the DTM module with informative features for cardiac activities, as shown in Figure~\ref{fig:HP_pipline}.

Lastly, the harmonic spectrogram with zero mask is added with the percussive spectrogram yielded from H$\&$P decomposition module to provide necessary information that preserves time consistency in the final Horcrux augmentation spectrogram as shown in Figure~\ref{fig:HP_pipline}. In addition, the added percussive spectrogram restricts the shift of data distribution brought by zero masks, and the augmented spectrogram still shares a similar representation with the normal spectrogram that is used in the evaluation or test stage of deep neural work training.

\begin{figure}[tb] 
    \centering 
    \includegraphics[width=0.9\columnwidth]{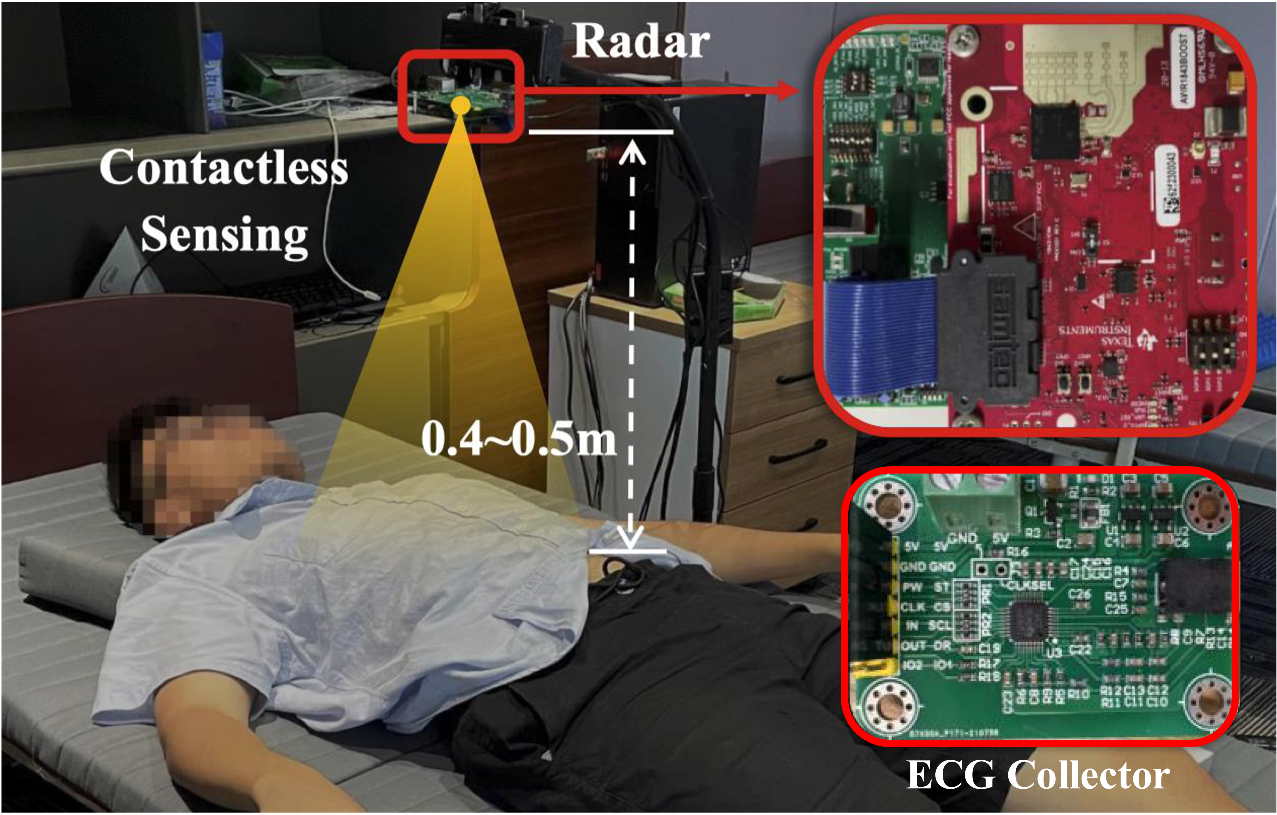}
    \caption{{Scenario for data collection~\cite{chen2022contactless}.}} 
    \label{fig:env_set} 
\end{figure}

\section{Experimental Result and Evaluation}
\subsection{Experimental Settings}
\subsubsection{Framework for Cardiac Feature Extraction}
To validate the proposed Horcrux on both classification and regression tasks in radar-based cardiac feature extraction, radarODE-MTL~\cite{zhang2024radarODEMTL} is selected as the only existing deep learning framework in the literature to simultaneously provide the ECG recovery and heartbeat detection result based on a multi-task learning paradigm. 

In general, radarODE-MTL receives the spectrogram as input and adopts a shared backbone with deformable 2D convolution layers~\cite{dai2017deformable} and ResNet structure~\cite{xue2024separable} to extract latent features from the spectrogram. Then, the task-specific decoders are implemented as:
\begin{itemize}
\item The ECG decoder is designed as several 1D transposed convolution blocks to recover the morphological features for ECG patterns, with root mean square error (RMSE) as the regression loss function.
\item The heartbeat decoder is designed as several 1D transposed convolution blocks followed by a linear projection module, and cross-entropy loss is adopted as a multi-class classification problem, i.e., each time index can be viewed as a possible class.
\end{itemize}

\subsubsection{Dataset Description}
The public dataset used in this research is from~\cite{chen2022contactless} with the radar signal collected by $77$GHz TI AWR-1843 radar with $3.8$GHz bandwidth, ensuring an ideal signal-to-noise ratio (SNR). The subject is asked to lie on the bed with the least body movement, and radar is set within a distance of $0.4-0.5$m right above the body as shown in Figure~\ref{fig:env_set}. Lastly, the ground ECG signal is collected by TI ADS1292 board with AC coupling and integrated right-leg drive (RLD) amplifier to remove potential baseline drift or power-line noise.

There are $91$ trails released by~\cite{chen2022contactless} with each trail containing radar and ground truth ECG signals for $3$ minutes and sampled at $200$Hz. In addition, several advanced beamforming or pre-processing techniques, such as 3D beamforming and dynamic signal clustering, are applied to amplify the interested cardiac features and remove the background or respiration noises. 

\subsubsection{Implementation Details}
The radarODE-MTL is coded with PyTorch and trained on the NVIDIA RTX A4000 with $16$GB memory. The training process lasts for $100$ epochs with batch size $22$, and the multi-task learning framework is optimized by eccentric gradient learning (EGA) algorithm and SGD optimizer with the hyperparameter $T=1$, learning rate $0.003$, weight decay $0.0004$, and momentum $0.937$~\cite{zhang2024radarODEMTL}. In addition, the width for the zero mask is empirically determined as $w_t=24$ and $w_f=12$, the time distance $\tau=0.5s$, and the dataset is formed as training, validation and testing sets following the portion of $80\%:10\%:10\%$. The input/ground truth data should be divided into the $4$-sec segments to adapt the design of radarODE-MTL. 
\begin{figure}[tb]
        \centering
        \subfloat[]{\label{fig:eg_0}\includegraphics[width=0.45\columnwidth]{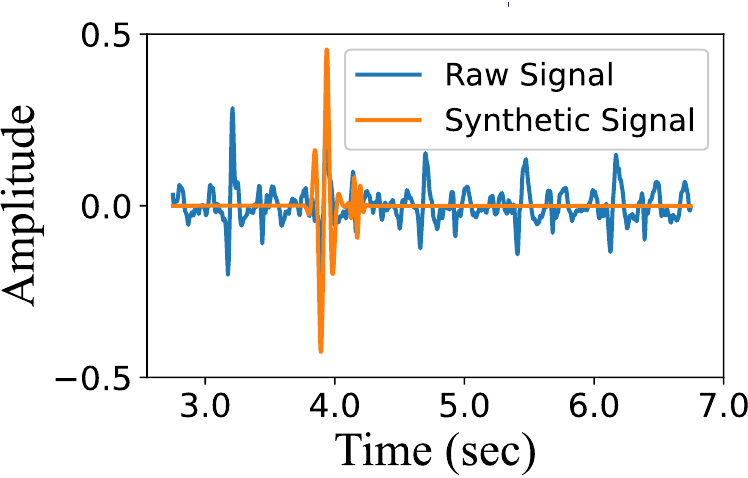}}
        \subfloat[]{\label{fig:eg_1}\includegraphics[width=0.45\columnwidth]{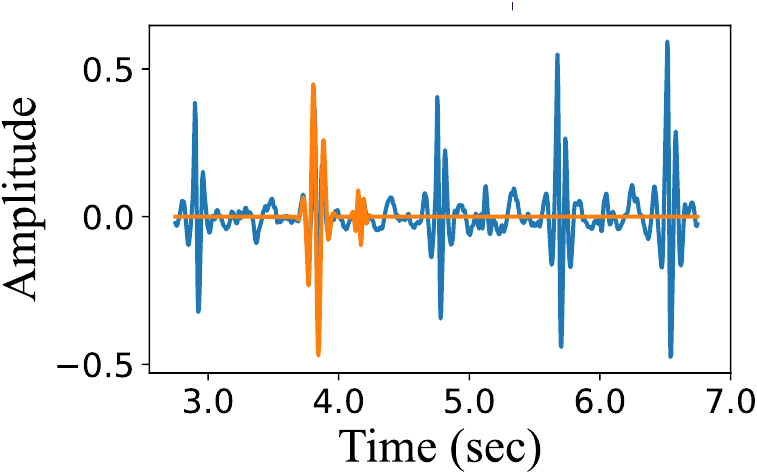}}\\
        \subfloat[]{\label{fig:eg_2}\includegraphics[width=0.45\columnwidth]{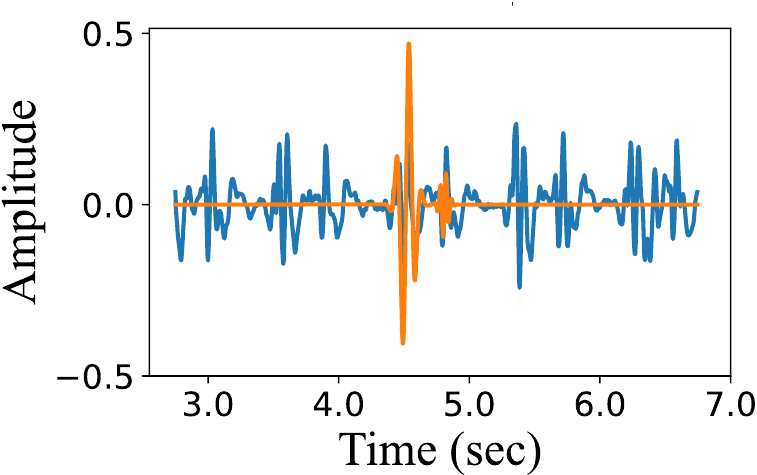}}
        \subfloat[]{\label{fig:eg_3}\includegraphics[width=0.45\columnwidth]{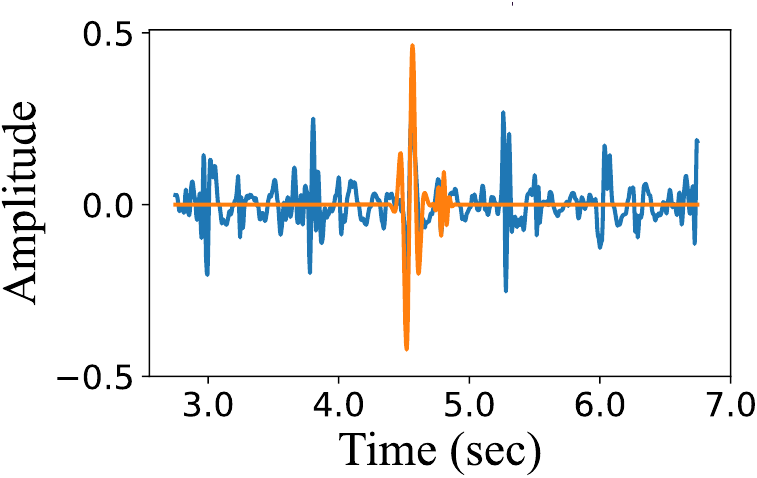}}\\
        \subfloat[]{\label{fig:eg_4}\includegraphics[width=0.45\columnwidth]{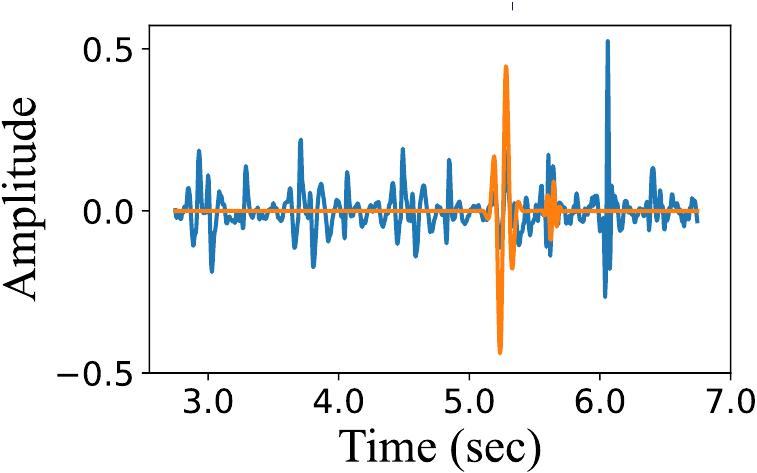}}
        \subfloat[]{\label{fig:eg_5}\includegraphics[width=0.45\columnwidth]{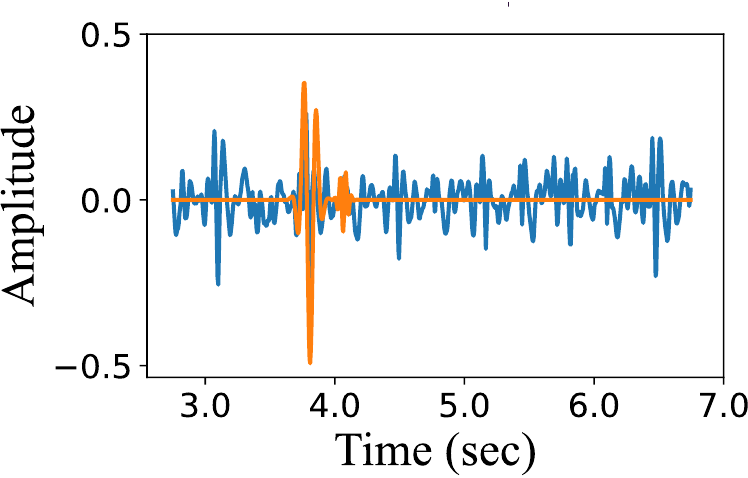}}
        \caption{Identified vibrations using DTM for radar signals with various qualities (illustrated in normalized amplitude).}
        \label{fig:temp_match}
\end{figure}
\subsubsection{Evaluation Metrics}
In this paper, the performance of the ECG recovery and heartbeat detection are evaluated from the following perspectives:
\begin{itemize}
\item \textbf{Root mean square error (RMSE)}: RMSE describes the morphological fidelity of the recovered ECG signal and is sensitive to the peak deviation.
\item \textbf{Pearson-correlation coefficient (PCC)}: PCC also reveals morphological fidelity but is sensitive to the general shape of the recovered ECG signal.
\item \textbf{Heartbeat error (H. E.)}: Heartbeat error simply shows the absolute timing error of the detected heartbeat with the ground truth.
\item \textbf{Missed detection rate (MDR)}: MDR is a metric to assess the performance of deep learning model against noises by calculating the percentage of the missed detected heartbeats, because the strong noise (e.g., body movement) or low SNR scenarios may drown the subtle cardiac features and cause missed detection.
\end{itemize}
In addition, to comprehensively evaluate the overall performance of each method, $\Delta m\%$ is adopted as in many multi-task learning studies to assess the performance based on multiple metrics as:
\begin{equation}\label{equ:mdr}
\Delta m\%=\frac{1}{n} \sum_{i=1}^{n} S_{i, j} \frac{M_{m,i}-M_{b, i}}{M_{b, i}} \times 100\%
\end{equation}
where $n$ is the number of metrics, $M_{m,i}$ means the performance of a method $m$ measured by metric $j$, $M_{b,i}$ means the performance of the baseline, and $S_{i, j}=1\text{ or }0$ if lower or higher values are better for the current metric (indicated by $\downarrow\text{ or }\uparrow$).

\subsection{Evaluation for DTM}
The DTM module aims to figure out the most prominent vibrations for a certain cardiac cycle within a $4$-sec input segment, and the illustration of the performance under different radar signal qualities (SNR) is shown in Figure~\ref{fig:temp_match}. It is worth noticing that the proposed DTM module not only identifies the prominent vibrations under good signal quality (e.g., Figure~\subref*{fig:eg_0} and~\subref*{fig:eg_2}), but could tolerant the low-SNR scenarios with certain constant noise drowning the characteristic peaks (e.g., Figure~\subref*{fig:eg_4} and~\subref*{fig:eg_5}). 

In addition, the DTM module could successfully identify $87\%$ and $53\%$ of the first and the second prominent vibrations ($v_1$ and $v_2$) within the absolute tolerance of $0.15$s~\cite{chen2022contactless}, ensuring a precise mask applied on the spectrogram in Horcrux. The specific impacts of the DTM module on the final ECG recovery quality will be evaluated in the later ablation study section, and the wrong detection in DTM will not degrade the overall performance of Horcrux because applying a random mask also contributes to the radar-based ECG recovery.

\begin{table}[t]
  \centering
  \caption{Comparison of Different Augmentation Methods}
  \begin{tabular}{lcccc|c}
  \toprule
  Methods & \makecell[c]{RMSE \\ (mV) } $\downarrow$ & PCC $\uparrow$ & \makecell[c]{H. E. \\ (ms) } $\downarrow$ & MDR $\downarrow$ &  $\Delta m\%$\\
  \midrule
  \textbf{Baseline} & 0.096 & 82.65\% & 8.82 & 6.73\% & 0.0\% \\
  \midrule
  C-Mixup~\cite{yao2022c} & \underline{0.088} & 82.25\% & 7.55 & 6.14\% & 7.80\% \\
  ADA~\cite{schneider2024anchor} & 0.097 & 81.28\% & 7.20 & 6.69\% & 4.12\% \\
  RC-Mixup~\cite{hwang2024rc} & 0.089 & 83.36\% & 7.72 & 5.79\% & 8.69\% \\
  \midrule
  \textbf{Horcrux} ($10\%$) & 0.096 & 82.24\% & 6.89 & 6.67\% & 5.63\% \\
  \textbf{Horcrux} ($15\%$) & 0.087 & \underline{84.75\%}& 6.96 & \textbf{5.19\%} & \underline{14.02\%} \\
  \textbf{Horcrux} ($20\%$) & \textbf{0.086} & \textbf{85.41\%} & \underline{6.21} & \underline{5.30\%} & \textbf{16.20\%} \\
  \textbf{Horcrux} ($25\%$) & 0.092 & 81.99\% & \textbf{6.16} & 6.36\% & 9.81\% \\
  \textbf{Horcrux} ($30\%$) & 0.094 & 80.18\% & 6.55 & 6.59\% & 6.78\% \\
  \bottomrule
  \multicolumn{6}{r}{The best/second best results are indicated by \textbf{Bold}/\underline{underline}.} \\
  \end{tabular}
  \label{tab:res}
\end{table}

\begin{table*}[t]
  \centering
  \caption{Ablation Study for Horcrux}
  \begin{tabular}{c|cc|cc|cccc|c}
  \toprule
  \multirow{3}*{\textbf{Input}} & \multicolumn{2}{c|}{\textbf{{Mask Type}}} & \multicolumn{2}{c|}{\textbf{{Selected Domain}}} & \multicolumn{4}{c|}{\textbf{\makecell[c]{Evaluation Metrics}}} & \multirow{3}*{$\Delta m\% $} \\
   & Random & DTM & Time & Frequency & \makecell[c]{RMSE \\ (mV) } $\downarrow$ & PCC $\uparrow$ & \makecell[c]{H. E. \\ (ms) } $\downarrow$ & MDR $\downarrow$ \\
  \midrule
  \multirow{7}*{\makecell[c]{Original \\ Spectrogram \\ (without Horcrux)}} &-&-&-&-& 0.096 & 82.65\% & 8.82 & 6.73\% & 0.0\% \\
    &\ding{52}&         &\ding{52}&          & 0.089 & 82.86\% & 7.14 & 7.38\% & 4.28\% \\
    &\ding{52}&         &         &\ding{52} & 0.088 & 83.53\% & 7.80 & 6.27\% & 7.04\% \\
    &\ding{52}&         &\ding{52}&\ding{52} & 0.087 & 83.03\% & 7.26 & 6.25\% & 8.42\% \\
    &         &\ding{52}&\ding{52}&          & 0.089 & 83.71\% & 9.53 & 6.34\% & -1.13\% \\
    &         &\ding{52}&         &\ding{52} & 0.099 & 80.00\% & 8.03 & 7.01\% & -0.24\% \\
    &         &\ding{52}&\ding{52}&\ding{52} & 0.098 & 81.19\% & 7.88 & 7.95\% & -2.41\% \\
  \midrule
  \multirow{7}*{\makecell[c]{H$\&$P \\ Spectrogram \\ (with Horcrux)}} &-&-&-&-& 0.091 & 83.73\% & 7.45 & 5.92\% & 8.57\% \\
    &\ding{52}&         &\ding{52}&          & \underline{0.084} & 85.40\% & \underline{6.90} & 6.71\% & 9.48\% \\
    &\ding{52}&         &         &\ding{52} & 0.086 & 84.03\% & 8.26 & \underline{5.35\%} & 9.86\% \\
    &\ding{52}&         &\ding{52}&\ding{52} & 0.085 & 84.89\% & 7.22 & 5.74\% & 11.80\% \\
    &         &\ding{52}&\ding{52}&          & \textbf{0.083} & 84.94\% & 7.04 & 5.72\% & \underline{13.18\%} \\
    &         &\ding{52}&         &\ding{52} & 0.085 & \textbf{87.20\%} & 7.96 & 5.66\% & 10.69\% \\
    &         &\ding{52}&\ding{52}&\ding{52} & 0.086 & \underline{85.41\%} & \textbf{6.21} & \textbf{5.30\%} & \textbf{16.20\%} \\
  \bottomrule
  \multicolumn{10}{r}{The best/second best results are indicated by \textbf{Bold}/\underline{underline}.} \\
  \end{tabular}
  \label{tab:aug-res}
\end{table*}
\subsection{Evaluation for Horcrux}
\subsubsection{Comparison with State-of-the-art Methods}
The performance of the proposed Horcrux is shown in Table~\ref{tab:res} in terms of RMSE, PCC, H. E. and MDR, with the comparison among the state-of-the-art augmentation methods for regressions tasks by mixing up the input and ground truth within homogeneous samples~\cite{hwang2024rc,schneider2024anchor,yao2022c}. In addition, the Horcrux is implemented on the dataset with different proportions to evaluate the impact of masks on the change of data distributions and model performance as shown in Table~\ref{tab:res}.

From Table~\ref{tab:res}, Horcrux achieves the best result when applying on $20\%$ of the dataset with an overall improvement of $16.20\%$ and outperforms all the existing augmentation methods, because the mix-up-based methods enhance the diversity of the dataset in sacrificing the time consistency inside one cardiac cycle. Therefore, the mix-up-based methods show less improvement or even degradation in RMSE and PCC, while Horcrux preserves the percussive component to maintain the time consistency inside the spectrogram. However, the masks applied on the spectrogram still change the data representation, and the overuse of Horcrux may cause the drift in the learned data representation or distribution, causing the degradation starting from the ECG recovery, as indicated by the bad RMSE and PCC in Table~\ref{tab:res} for Horcrux ($25\%$ and $30\%$).

\subsubsection{Ablation Study}
The improved performance achieved by Horcrux comes from two aspects, i.e., H$\&$P decomposition and DTM module, and an ablation study is performed to specify the contributions as shown in Table~\ref{tab:aug-res}. Firstly, the original spectrogram is used as inputs with masks applied in the time or frequency domain randomly or based on the DTM module. It is worth noticing that simply applying random masks on the original set could achieve an improvement of $8.42\%$, and masking the frequency domain ($7.04\%$) contributes more than masking the time domain ($4.28\%$), because zero mask will cover all the useful information and affects heartbeat detection. This phenomenon can also be verified by precisely masking the prominent vibrations based on DTM, causing the heavy degradation with negative $\Delta m\%$ as shown in Table~\ref{tab:aug-res}.

In addition, the H$\&$P decomposition itself also improves the cardiac features extraction with an improvement of $8.57\%$, because the percussive and harmonic features are both enhanced after adding them up. Afterward, the random mask could further improve the performance to $9.86\%$, while the DTM-based mask achieves the largest improvement of $16.20\%$. It is also worth noticing that only masking the time domain ($13.18\%$) achieves a better result than covering the frequency domain ($10.69\%$), because the H$\&$P decomposition preserves the percussive information, and the heartbeat is still detectable even after masking the prominent vibrations precisely.
\begin{figure}[tb] 
    \centering 
    \includegraphics[width=0.9\columnwidth]{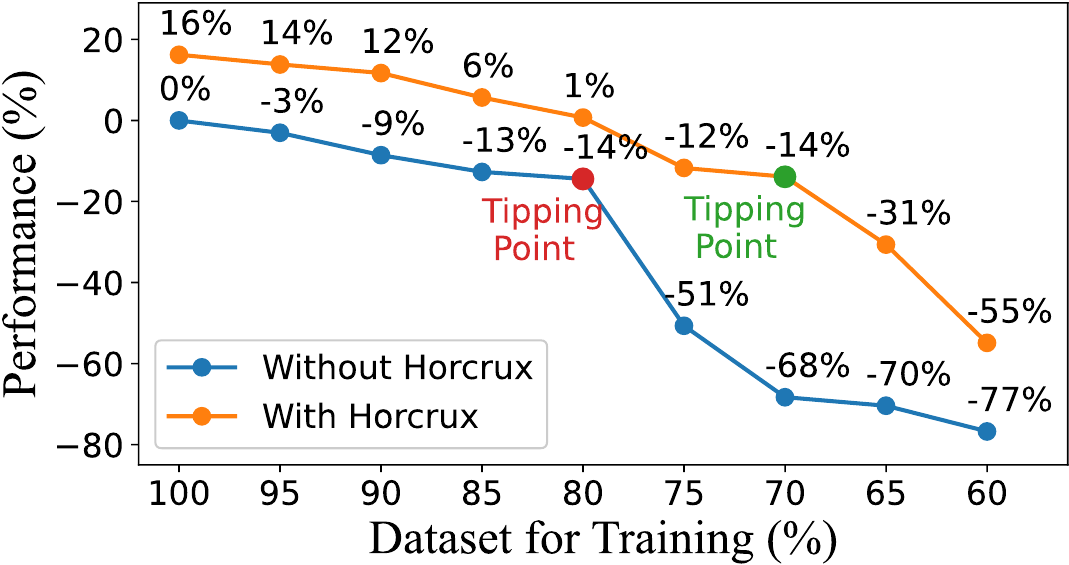}
    \caption{Performance of using different scales of the dataset.} 
    \label{fig:perfor} 
\end{figure}

\subsubsection{Performance under Different Dataset Scale}
As a data augmentation technique, Horcrux is tested with different scales of dataset with the overall performance $\Delta m\%$ shown in Figure~\ref{fig:perfor}. It is notable that the degradation of the performance is not in a linear relationship with the input data scale, showing an obvious tipping point after axing the input data to $80\%$ without augmentation. After applying Horcrux, the performance is boosted using the same scale of input data as already proved by the previous ablation study. In addition, the deep learning model with Horcrux reveals a mild degradation with reduced input data, and the occurrence of the tipping point is also postponed to $70\%$ as shown in Figure~\ref{fig:perfor}, because the faithful alteration of the data representation could improve the generalization of deep learning model and alleviate the risk of overfitting.

\section{Conclusions}
This study proposes the data augmentation technique, Horcrux, for radar-based cardiac feature extraction with deep learning frameworks. Different from the traditional augmentation methods that may block the crucial parts or destroy the time consistency between input and ground truth, Horcrux preserves the percussive features of the input signal and only masks the cardiac-feature-related region on the harmonic spectrogram, enhancing the diversity of the limited dataset and also encouraging the deep learning model to only focus on the domain-preserved features. Compared with the existing methods, Horcrux reveals outstanding performance in heartbeat detection and ECG recovery and could effectively alleviate the degradation caused by the limited data. In the future, Horcrux has the potential to be extended to other spectrogram-based tasks and needs to be further evaluated for certain downstream tasks such as radar-based cardiovascular disease diagnosis. 
\bibliographystyle{IEEEtran}
\bibliography{reference}
\addcontentsline{toc}{section}{Reference}
\normalsize

\vspace{12pt}

\end{document}